\batchmode
\makeatletter
\def\input@path{{/Users/ananyabalakrishna/Documents/Disemmination/Writing/Arxived/To_be_arxived/2_Scaling_surface_energy/Lyx/}}
\makeatother
\documentclass[10pt,twocolumn]{article}
\usepackage[latin9]{inputenc}
\usepackage[letterpaper]{geometry}
\usepackage{color}
\usepackage{amsmath}
\usepackage{amssymb}
\usepackage{graphicx}
\usepackage{setspace}
\usepackage[unicode=true,pdfusetitle,
 bookmarks=true,bookmarksnumbered=false,bookmarksopen=false,
 breaklinks=false,pdfborder={0 0 1},backref=false,colorlinks=false]
 {hyperref}

\makeatletter

\newcommand{\lyxmathsym}[1]{\ifmmode\begingroup\def\b@ld{bold}
  \text{\ifx\math@version\b@ld\bfseries\fi#1}\endgroup\else#1\fi}

\usepackage{ragged2e}
\usepackage{booktabs}
\usepackage{authblk}
\usepackage{setspace}
\usepackage{titlesec}
\titlespacing*{\section}{0pt}{1.1\baselineskip}{\baselineskip}

\usepackage[parfill]{parskip}
\usepackage{graphicx}
\usepackage{cite}
\usepackage{notoccite}

\title{Scale effects and the formation of polarization vortices in tetragonal ferroelectrics}
\author{Ananya Renuka Balakrishna, John E. Huber}
\affil{Department of Engineering Science, University of Oxford, Oxford, OX1 3PJ, United Kingdom}

\date{}                    %% if you don't need date to appear

\makeatother

\begin{document}
\twocolumn[   
\begin{@twocolumnfalse}  

\maketitle 
\begin{abstract}
\begin{singlespace}
\textcolor{black}{Vortices consisting of $90^{\circ}$ quadrant domains
are rarely observed in ferroelectrics. Although experiments show polarization
flux closures with stripe domains, it is as yet unclear why pure single
vortices are not commonly observed. Here we model and explore the
energy of polarization patterns with vortex and stripe domains, formed
on the square cross-section of a barium titanate nanowire. Using phase-field
simulations, we calculate the associated energy of polarization patterns
as a function of nanowire width. Further, we demonstrate the effects
of surface energy and electrical boundary conditions on equilibrium
polarization patterns. The minimum energy equilibrium polarization
pattern for each combination of surface energy and nanowire width
is mapped for both open-circuit and short-circuit boundary conditions.
The results indicate a narrow range of conditions where single vortices
are energetically favorable: nanowire widths less than about $30\mathrm{nm}$,
open-circuit boundary condition, and surface energy of less than 4N/m.
Short-circuit boundary conditions tend to favor the formation of a
monodomain, while surface energy greater than 4N/m can lead to the
formation of complex domain patterns or loss of ferroelectricity.
The length scale at which a polarization vortex is energetically favorable
is smaller than the typical size of nanoparticle in recent experimental
studies. The present work provides insight into the effects of scaling,
surface energy and electrical boundary conditions on the formation
of polarization patterns.}
\end{singlespace}
\end{abstract}
\end{@twocolumnfalse} ]
\begin{singlespace}

\section*{{\small{}Introduction}}
\end{singlespace}

\begin{singlespace}
\noindent \textcolor{black}{\small{}Although vortices of magnetic
domains have been experimentally observed in ferromagnets \cite{key-1,key-2,key-3,key-4,key-5},
it is intriguing to note that the analogous simple polarization vortices
are not seen in ferroelectrics. Models such as the time-dependent
Ginzburg-Landau theory suggest that polarization vortices should form
under certain conditions \cite{key-6,key-7,key-8,key-9}.}{\small \par}
\end{singlespace}

\noindent \textcolor{black}{\small{}Polarization vortices possess
tremendous potential for the design of nanoscale devices such as memory
elements \cite{key-10,key-11,key-12,key-13} and transducers \cite{key-6,key-14,key-15}.
With the progressive miniaturization of electronics, the functional
properties of ferroelectric domain patterns are of increasing importance
\cite{key-16,key-17,key-18,key-19,key-20}. Hence much current research
is directed towards finding polarization vortices and studying their
nanoscale properties in detail \cite{key-21,key-22,key-23,key-24,key-25,key-26,key-27,key-28,key-29,key-30,key-31}.}{\small \par}

\noindent \textcolor{black}{\small{}Polarization flux closures in
the form of bundles of $90^{\circ}$ stripe domains oriented to form
a vortex have been imaged by McGilly and Gregg in PbZr$_{(0.42)}$Ti$_{(0.58)}$O$_{3}$
nanodots \cite{key-25}. Similar stable flux closures have been observed
by McQuaid et al. in BaTiO$_{3}$\cite{key-28}. However, these flux-closures
consist of $90^{\circ}$ stripe domains \cite{key-23,key-30} and
so differ from the classic polarization vortex consisting of $90^{\circ}$
quadrant domains, which is well-known in ferromagnetic materials \cite{key-32,key-33}.
Polarization vortex patterns consisting of dipole flux closures \cite{key-26,key-28}
or quadrupole chains \cite{key-26}, have been observed. However,
these vortices have $180^{\circ}$ domain walls at their core and
so differ from the classic polarization vortex in small particles
as predicted by Kittel \cite{key-1}. Although, the direct observation
of polarization rotation which facilitates the formation of a vortex
has been established \cite{key-24}, the classic vortex polarization
pattern continues to be elusive \cite{key-32}. This leads us to consider
the question \textendash{} Why are these polarization vortices not
seen in experiments? }{\small \par}

\noindent \textcolor{black}{\small{}In the present work, a BaTiO$_{3}$
nanowire in the tetragonal phase, with square cross-section is modelled
in isothermal conditions. The wire is assumed to extend indefinitely
out of the model plane, such that plane strain and plane electric
field conditions apply. Minimum cross-sectional widths of 20nm were
considered, noting that ferroelectricity has been observed in BaTiO$_{3}$
structures from a few nanometers in size upwards \cite{key-34,key-35,key-36}.
The wire is assumed to be simply supported, with traction free surfaces.
We consider two distinct electrical boundary conditions: open-circuit
and short-circuit, by applying zero normal component of electric displacement
and zero voltage boundary conditions, respectively. }{\small \par}

\noindent \textcolor{black}{\small{}A phase-field model previously
developed by Landis and co-workers \cite{key-37,key-38,key-39} calibrated
for BaTiO$_{3}$ is used to find equilibrium states. This model has
been applied as a design tool \cite{key-6,key-11} and to study domain
wall interactions in ferroelectrics \cite{key-37,key-38}. The model
describes the Helmholtz free energy, $\psi$ as a function of polarization,
$P_{i}$, which is the order parameter, strain, $\epsilon_{ij}$ and
electric displacement, $D_{i}$ \cite{key-37}:
\begin{align}
\psi & =\psi_{g}+\psi_{d}\label{eq:1}\\
\psi_{g} & =\frac{1}{2}a_{ijkl}P_{i,j}P_{k,l}\label{eq:2}
\end{align}
}\textcolor{black}{\footnotesize{}
\begin{align}
\psi_{d} & =\frac{1}{2}\overline{a}_{ij}P_{i}P_{j}+\frac{1}{4}\overline{\overline{a}}_{ijkl}P_{i}P_{j}P_{k}P_{l}+\frac{1}{6}\overline{\overline{\overline{a}}}_{ijklmn}P_{i}P_{j}P_{k}P_{l}P_{m}P_{n}\nonumber \\
 & ~+\frac{1}{8}\overline{\overline{\overline{\overline{a}}}}_{ijklmnrs}P_{i}P_{j}P_{k}P_{l}P_{m}P_{n}P_{r}P_{s}+b_{ijkl}\epsilon_{ij}P_{k}P_{l}+\frac{1}{2}c_{ijkl}\epsilon_{ij}\epsilon_{kl}\nonumber \\
 & ~+\frac{1}{2}f_{ijklmn}\epsilon_{ij}\epsilon_{kl}P_{m}P_{n}+\frac{1}{2}g_{ijklmn}\epsilon_{ij}P_{k}P_{l}P_{m}P_{n}\nonumber \\
 & \ +\frac{1}{2\kappa_{0}}(D_{i}-P_{i})(D_{i}-P_{i})\label{eq:3}
\end{align}
}\textcolor{black}{\small{}The form of Eq. \ref{eq:1} \textendash{}
\ref{eq:3} is identical to that in the work of Landis and co-workers
\cite{key-37,key-38,key-39}, where the detailed meaning of the specific
terms and material properties are explained. For our purposes, we
note that the gradient energy, $\psi_{g}$, includes energy due to
polarization variation at domain walls or surfaces. The domain energy, $\psi_{d}$ ,
accounts for the elastic, piezoelectric and dielectric energy due
to distortion away from the spontaneously polarized state. The domain
evolution follows a generalized Ginzburg-Landau equation \cite{key-37}:
\begin{equation}
\left(\frac{\partial\psi}{\partial P_{i,j}}\right)_{,j}-\frac{\partial\psi}{\partial P_{i}}=\beta P_{i},\label{eq:4}
\end{equation}
}{\small \par}

\noindent \textcolor{black}{\small{}where $\beta$ is the polarization
viscosity, which was controlled as a relaxation parameter in the simulation
to allow equilibrium states $\beta=0$ with to be found. In the simulation,
the polarization vector is constrained to lie in-plane with no other
boundary conditions applied; it is governed by Eq. \ref{eq:4}. }{\small \par}

\noindent \textcolor{black}{\small{}The phase-field model is solved
using finite element methods with the element size chosen such that
a $180^{\circ}$ domain wall spans four elements. The nodal displacements
enable the strain and polarization gradient to be computed; hence
the energy density was found from Eq. \ref{eq:1} \textendash{} \ref{eq:3}. }{\small \par}

\begin{figure}
\begin{centering}
{\footnotesize{}\includegraphics[width=0.8\columnwidth]{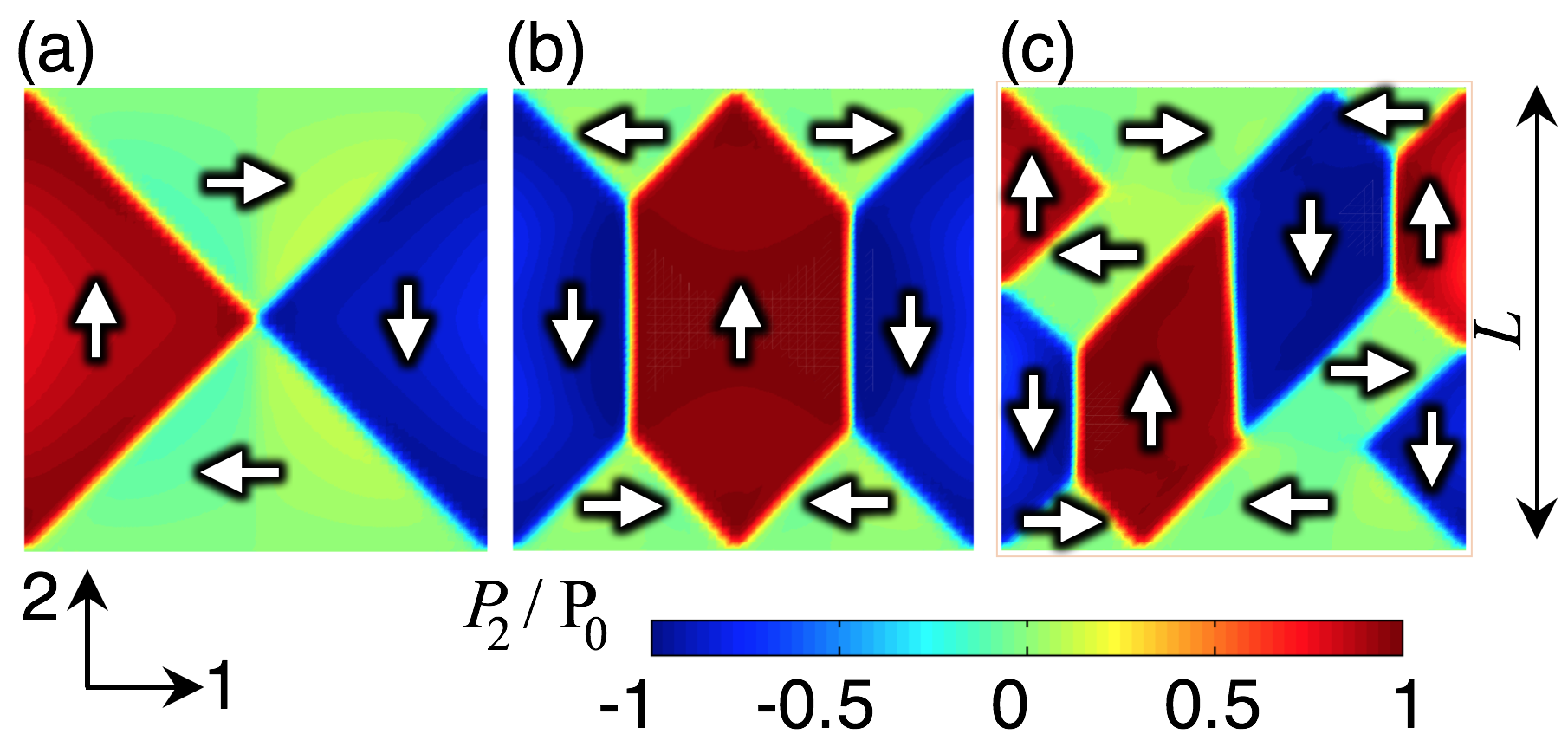}}
\par\end{centering}{\footnotesize \par}
{\footnotesize{}\caption{{\small{}\label{fig:1}Polarization, $P_{2}$, of the three patterns
explored (a) Type I (b) Type II (c) Type III. $P_{0}=0.26\mathrm{C/m^{2}}$;
$L=40\mathrm{nm}$.}}
}{\footnotesize \par}
\end{figure}

\noindent \textcolor{black}{\small{}To identify commonly occurring
domain arrangements, the initial state of the model was set by assigning
random polarization values, $\frac{-P_{0}}{2}\leq P_{i}\leq\frac{P_{0}}{2}$,
at each node. Wire cross-sections in the size range $20\mathrm{nm}\leq L\leq40\mathrm{nm}$
were simulated. From this random starting state the simulation converged
on equilibrium states, which frequently resulted in one of the three
types of domain arrangement shown in Fig. \ref{fig:1}(a\textendash c);
see also Xue et al. \cite{key-40} where similar structures were found
using Monte Carlo methods. Type I {[}Fig. \ref{fig:1}a{]} consists
of a classic vortex while Type II is identified with two vortices
and stripe domains {[}Fig. \ref{fig:1}b{]}. This structure is closely
related to the double-closure pattern with domain wall vertices as
observed by McQuaid et al. \cite{key-29}. Type III is a more complicated
pattern of domains forming a flux closure {[}Fig. \ref{fig:1}c{]}. }{\small \par}

{\footnotesize{}}
{\footnotesize \par}

\noindent \textcolor{black}{\small{}Having established three types
of polarization patterns with vortex and stripe domains {[}Fig. \ref{fig:1}(a\textendash c){]}
that can typically form on the cross-section of a nanowire, these
patterns were studied further to establish the size dependency of
their stability. Nanowire cross-sections of size 20\textendash 80nm
were simulated with initial conditions that forced each of the three
patterns of Fig. \ref{fig:1} to form. This was achieved by initializing
the simulation with the polarization at each node set to match one
of the domain patterns in Fig. \ref{fig:1}, while the nodal displacements
and electric potential were initialized at zero. If the simulations
reached equilibrium without pattern change, this indicated the stability
of the pattern and the associated energy was thus found as a function
of size. As the size was increased, the type I pattern remained stable,
but other patterns became energetically favorable. }{\small \par}

\begin{figure}
\begin{centering}
\includegraphics[width=0.8\columnwidth]{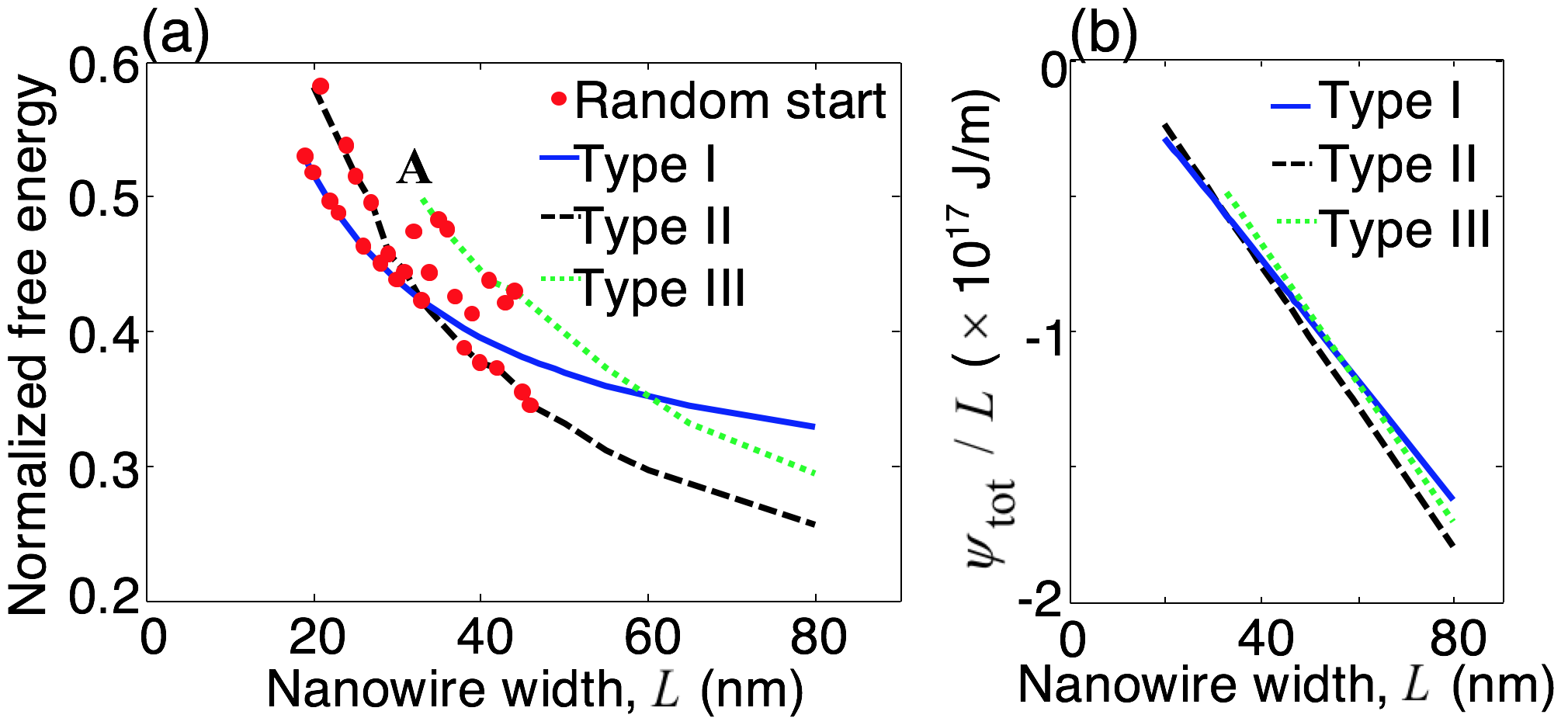}
\par\end{centering}
\caption{{\small{}\label{fig:2}(a) Normalized free energy, $(\psi_{0}-\psi)/\psi_{0}$,
values as a function of nanowire width for three types of polarization
patterns obtained from phase-field simulations. (b) Total free energy
per unit width, $\psi_{\mathrm{tot}}/L$, versus $L$, showing best
fit straight lines. }}
\end{figure}

\noindent \textcolor{black}{\small{}The resulting free energy for
each of the three types of polarization patterns obtained from the
phase-field simulations is shown in Fig. \ref{fig:2}a as a function
of nanowire width, $L$. The energy per unit volume, $\psi$, is normalized
as $(\psi_{0}-\psi)/\psi_{0}$, \cite{key-11} where $\psi_{0}$ corresponds
to the energy per unit volume of a monodomain element in a spontaneously
polarized state. The energy curves of type I and type II cross at
$L=34\mathrm{nm}$ indicating a dependence of minimum energy state
upon nanowire width. The type III pattern is not stable for $L<35\mathrm{nm}$:
even if the simulation is started with polarization matching the type
III pattern, other flux closures with lower energy form. The lower
size limit for stability of the type III pattern is indicated by \textquotedblleft A\textquotedblright{}
in Fig. \ref{fig:2}a. Also shown in Fig. \ref{fig:2}a are the results
obtained from starting the simulations with randomly polarized states.
These data jump back and forth between the main three types of polarization
pattern indicating that the stable state found is highly dependent
on the starting conditions. }{\small \par}

\noindent \textcolor{black}{\small{}Since the model size is much larger
than the intrinsic length scale due to domain wall width, the total
energy in volume $V$ due to polarization gradient, $\int\psi_{g}\mathrm{d}V$,
scales approximately with domain wall area whereas domain energy,
$\int\psi_{d}\mathrm{d}V$, scales with the volume of polarized domains.
Then, defining the total free energy $\psi_{\mathrm{tot}}=\int\psi\mathrm{d}V$,
and defining the volume to have an out-of-plane depth $D$, the energy
of a given pattern is: }{\small \par}

\noindent \textcolor{black}{\small{} 
\begin{equation}
\psi_{\mathrm{tot}}=aLD+bL^{2}D\label{eq:5}
\end{equation}
}{\small \par}

\noindent \textcolor{black}{\small{}where, the coefficients $a$ and
$b$ for each pattern are estimated using linear regression of the
data in Fig. \ref{fig:2}a, {[}see Fig. \ref{fig:2}b{]}. }{\small \par}

\noindent \textcolor{black}{\small{}For a given nanowire width, $L$,
the gradient energy, $\psi_{g}$ of the three types of polarization
patterns is governed by coefficient $a(\times10^{16}\mathrm{J/m^{2}})$
which is related by type III \textgreater{} type II \textgreater{} type I.
While $\psi_{d}$ is governed by coefficient $b(\times10^{24}\mathrm{J/m^{3}})$
and follows the reverse order type I \textgreater{} type II \textgreater{} type III.
Thus the multidomain patterns reduce their domain energy at the cost
of increased gradient energy. }{\small \par}

\noindent \textcolor{black}{\small{}When $L<34\mathrm{nm}$, the percentage
contribution of $\psi_{g}$ to $\psi$ is significant $(25\lyxmathsym{\textendash}30\%)$,
causing polarization patterns with greater domain wall area to possess
greater energy {[}Fig. \ref{fig:2}a{]}. This makes the classic polarization
vortex (type I) energetically favorable when $L<34\mathrm{nm}$. However,
for $L>50\mathrm{nm}$, $\psi_{d}$ dominates the energy. Thus, for
$L>50\mathrm{nm}$, polarization patterns with stripe domains become
favorable. The balance of energy contributions: $\psi_{g}$, $\psi_{d}$
and nanowire width, $L$ determines the minimum energy polarization
pattern {[}Fig. \ref{fig:2}(a\textendash b), Eq. \ref{eq:5}{]}.
Noting that the energy is well fitted by  $\frac{\psi_{\mathrm{tot}}}{L^{2}D}=\frac{a}{L}+b$,
then in the limit as $L$ becomes large, $\frac{\psi_{\mathrm{tot}}}{L^{2}D}\rightarrow b$.
Hence the curves in Fig. \ref{fig:2}a asymptotically approach a constant
energy per unit volume. This suggests that the type III domain pattern
will become favorable at larger scales. However in the present study
calculations did not go beyond $L=80\mathrm{nm}$; it is likely that
other low energy patterns will become favorable before the cross-over
from type II to type III is reached. }{\small \par}

\begin{figure}
\begin{centering}
{\small{}\includegraphics[width=0.8\columnwidth]{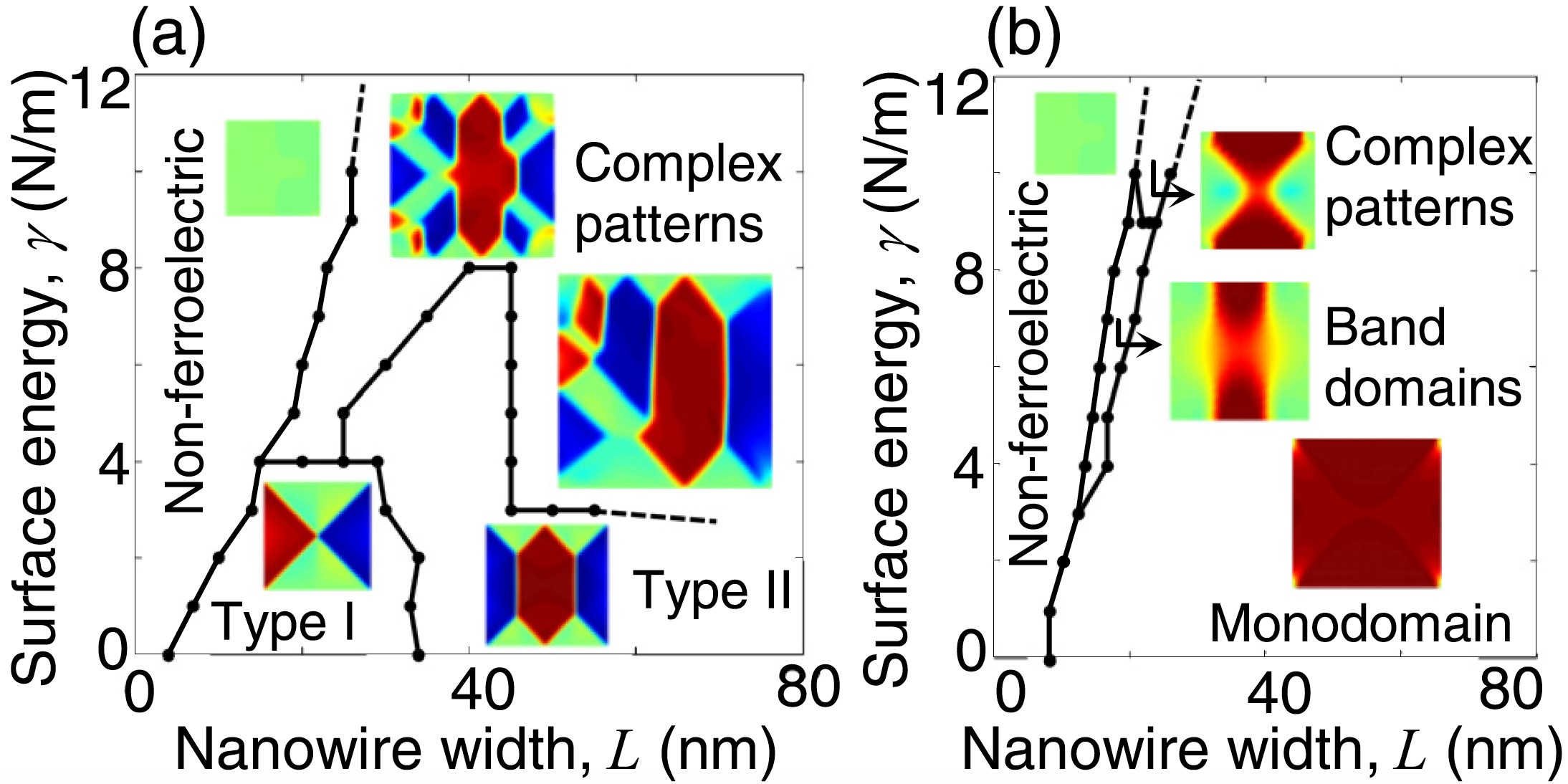}}
\par\end{centering}{\small \par}
{\small{}\caption{{\small{}\label{fig:3}Phase diagrams showing minimum energy domain
patterns with (a) open-circuit boundary conditions (b) short-circuit
boundary conditions. Inset diagrams show typical patterns of polarization
$P_{2}/P_{0}$.}}
}{\small \par}
\end{figure}

\noindent \textcolor{black}{\small{}Up to this point, surface energy
was neglected and only open-circuit boundary conditons were considered.
However, at the nanoscale, both surface energy, $\gamma$, \cite{key-41,key-42,key-43}
and electrical boundary conditions \cite{key-36,key-44} affect the
formation of ferroelectric domains. For BaTiO$_{3}$ nanowires, experiments
and theoretical considerations suggest $\gamma$ values of about 0.68N/m,
\cite{key-36,key-44} however the presence of depolarization field
and surface layer effects can cause local variation in $\gamma$ values
and have led to greater estimates of $\gamma$, around 10N/m \cite{key-44}.
Other authors found values within this range depending on shape and
surface conditions including chemical environment \cite{key-42,key-43}.
Hence, we allow surface energy values in the range $0\mathrm{N/m\leq}\gamma\leq10\mathrm{N/m}$
in the model. Morozovska et al. \cite{key-45,key-46,key-47}, have
modelled the effect of a surface tension proportional to local curvature
of cylindrical nanoparticles and nano-rods, via the free energy function.
For our case involving a square cross-section nanowire we approximate
the effect of surface tension by applying surface force boundary conditions
at corners only, neglecting second order effects on surface curvature
due to deformation. The surface effect on the Helmholtz free energy
$\psi$ in a narrow region near the nanowire surface is also neglected.
Local polarization orientation relative to the surface also affects
the value of $\gamma$, and indeed the relation between the surface
energy and the resulting surface stresses is expected to be anisotropic.
However since this study focuses on flux closures, we expect polarization
to be parallel to the surface and so neglect this effect. We further
consider short-circuit or open-circuit boundary conditions. A \textquotedblleft phase
diagram\textquotedblright{} mapping the minimum energy equilibrium
state for each combination of $\gamma$ and $L$, with open-circuit
and short-circuit boundary conditions is shown in Fig. \ref{fig:3}(a\textendash b).
Boundaries on the diagram indicate approximately the location of points
where the patterns associated with the adjacent regions have equal
energy; markers show specific points calculated on each boundary. }{\small \par}

\noindent \textcolor{black}{\small{}In nanowire cross-section with
open-circuit boundary conditions {[}Fig. \ref{fig:3}a{]}, type I
polarization pattern is the minimum energy arrangement in a region
with $10\mathrm{nm}<L<30\mathrm{nm}$ approximately, and surface energy,
$\gamma<4\mathrm{N/m}$. At greater values of surface energy, complex
patterns with multiple domains are favored, while nanowire widths
$L>30\mathrm{nm}$ favor the type II pattern as the lowest energy
state. BaTiO$_{3}$ nanowires with open-circuit boundary condition
are non-ferroelectric for combinations of high values of surface energy,
$\gamma>4\mathrm{N/m}$ and low values of width, $L<10\mathrm{nm}$.
This is manifested in the model by the disappearance of tetragonality
$(\epsilon_{11}=\epsilon_{22}=0)$ and polarization $P_{1}=P_{2}=0$.
By contrast, in a nanowire with short-circuit boundary condition {[}Fig. \ref{fig:3}b{]},
a monodomain state is favored at low values of surface energy $\gamma<2\mathrm{N/m}$,
when $L>2\mathrm{nm}$. There is a narrow region where polarization
patterns with band-like domains and complex patterns with multiple
domains are observed, with $10\mathrm{nm}<L<20\mathrm{nm}$ and $\gamma>4\mathrm{N/m}$.
Finally, the model suggests that BaTiO$_{3}$ nanowires with short-circuit
boundary condition are non-ferroelectric when $\frac{\gamma}{L}>8.5\times10^{8}\mathrm{N/m^{2}}$.}{\small \par}

\noindent \textcolor{black}{\small{}The phase diagrams in Fig. \ref{fig:3}
are consistent with several aspects of experimental observations.
The lack of experimental observations of the classic (type I) polarization
vortex is explained by two factors. First, scale effects are important
in that typical experiments which map in-plane polarization patterns
use sample sizes of order 100nm upwards \cite{key-25,key-27,key-30,key-32}.
The simulations suggest that the type I vortex is a high energy state
at this scale. Second, surface environments in experiments often include
polar or ionic species that may act as charge carriers, providing
some conductivity \cite{key-21,key-30,key-36,key-48,key-49}. Again,
the simulations suggest that the type I vortex is unlikely to form
in conductive environments. Other features that agree with experiment
include the loss of tetragonality at small scales; this has been observed
in barium titanate nanowires at scales of 10nm or less \cite{key-50,key-51,key-52}.
Meanwhile in nanowires with short-circuit boundary conditions, ferroelectricity
has been observed down to the nanometer scale provided the surface
energy is low, consistent with Fig. \ref{fig:3}b \cite{key-36}.
At greater length scales, 80nm upward, complex domain patterns including
several or many domains are typical \cite{key-25,key-27,key-32,key-48}. }{\small \par}

\noindent \textcolor{black}{\small{}In conclusion, we used a phase-field
simulation to study the effects of scale and surface conditions on
the polarization patterns that can form in BaTiO$_{3}$ nanowires.
There exists a narrow range of scale and surface conditions for which
the classic single polarization vortex is likely to form. The study
thus provides an insight into the absence of experimental observation
of classic polarization vortices of the form described by Kittel:
typical experiments in nanowires and nanoparticles do not operate
in the regime where such vortices are energetically favorable. At
scales on the order of a hundred nanometers, the classic single vortex
is unlikely to appear because of high domain energy, which is lowered
in multiple domains. At smaller scales, ferroelectricity is plagued
by surface energy and surface conductance that affect the formation
of a single vortex, making it elusive. }{\small \par}

\noindent \textcolor{black}{\small{}The authors wish to thank Prof.
C. M. Landis and Dr. D. Carka for help in providing program codes
and advice. We also acknowledge Dr. I. Münch for advice on modelling.
A. Renuka Balakrishna gratefully acknowledges support of the Felix
scholarship trust.}{\small \par}

\end{document}